%% file: paper.tex
\input{header-1-packages}

\input{header-2-macros}

\input{header-3-acronyms}

\input{header-3a-editing}
\begin{document}

\input{header-4-title+authors}

\date{\today}

\input{header-5-abstract}

\maketitle

\input{sec-1}

\input{sec-3}

\input{sec-4}

\input{acknowledgments}

\bibliographystyle{apsrev4-2} 
\showtitleinbib
\bibliography{fixapsbib,refs,smg,GNref}
\onecolumngrid\vspace*{2\baselineskip}\twocolumngrid
\end{document}

%% file: header-2-macros.tex
\newcommand{\ungroup}[1]{#1}
\newcommand{\withbreak}[1]{\expandafter\ungroup#1}

\def\dounbracket[#1]{#1}

\newcommand{\SU}{{\mathrm{SU}}}

\newcommand{\ubar}{{\bar{u}}}
\newcommand{\dbar}{{\bar{d}}}

\renewcommand\vec\mathbf

\newcommand\redsout{\bgroup\markoverwith{\textcolor{red}{\rule[0.5ex]{2pt}{1.4pt}}}\ULon}

\newcommand\dootimesall[2]{\ifx0#1\else\mathbf{#1}\ifx0#2\else\def\mytmp{\otimes\dootimesall{#2}}\expandafter\expandafter\expandafter\mytmp\fi\fi}

%% file: header-3-acronyms.tex
\usepackage{relsize}

\usepackage[acronyms, nohypertypes={acronym}, nopostdot, style=super, nonumberlist, toc]{glossaries}
\setacronymstyle{long-sc-short}
\glsenableentrycount

\newacronym{PNA}{pna}{particle-number-algorithm}
\newacronym{SFA}{sfa}{spin-flip-algorithm}

\newacronym{WF}{wf}{Wilson-Fisher}
\newacronym{AF}{af}{asymptotically free}

\newacronym{RG}{rg}{renormalization group}

\newacronym{QIS}{qis}{Quantum Information Science}
\newacronym{PPT}{ppt}{positive-semidefinite partial transpose}
\newacronym{KS}{ks}{Kogut-Susskind}
\newacronym{NPT}{npt}{negative partial transpose}
\newacronym{SMG}{smg}{symmetric mass generation}
\newacronym{SMF}{smf}{symmetric massive fermion}
\newacronym{MF}{mf}{massless fermion}
\newacronym{SSB}{ssb}{spontaneous symmetry breaking}

\newacronym{AS}{as}{Anti-Symmetric}

\newacronym[longplural={conformal field theories}]{CFT}{cft}{conformal field theory}
\newacronym[longplural={lattice field theories}]{LFT}{lft}{lattice field theory}
\newacronym[longplural={effective field theories}]{EFT}{eft}{effective field theory}
\newacronym[longplural={quantum field theories}]{QFT}{qft}{quantum field theory}
\newacronym[longplural={lattice gauge theories}]{LGT}{lgt}{lattice gauge theory}
\newacronym[longplural={monomer-dimer tensor-networks}]{MDTN}{mdtn}{monomer-dimer tensor-network}

\newacronym[]{DMRG}{dmrg}{Density Matrix Renormalization Group}
\newacronym[]{TFIM}{tfim}{Transverse Field Ising Model}

\newacronym[]{LOCC}{locc}{Local Operations and Classical Communicaton}
\newacronym[]{OBC}{obc}{open boundary conditions}

\newacronym{MPS}{mps}{matrix product states}

\newacronym{JLP}{jlp}{Jordan-Lee-Preskill}

\newacronym{BBN}{bbn}{big bang nucleosynthesis}

\newacronym{LEC}{lec}{low-energy constant}

\newacronym{QCD}{qcd}{quantum chromodynamics}
\newacronym{MC}{mc}{Monte Carlo}

\newacronym{IR}{ir}{infrared}
\newacronym{UV}{uv}{ultraviolet}

\newacronym{QED}{qed}{quantum electrodynamics}
\newacronym{SNR}{snr}{signal-to-noise ratio}

\newacronym{NLSM}{nlsm}{nonlinear sigma model}

\newacronym{CL}{cl}{Complex Langevin}

\newacronym{CSA}{csa}{Cartan subalgebra}

\newacronym{AFQMC}{afqmc}{auxiliary field quantum Monte Carlo}
\newacronym{iHMC}{ihmc}{imaginary-mass Hybrid Monte Carlo}

\newacronym{MCMC}{mcmc}{Markov Chain Monte Carlo}

\newacronym{QI}{qi}{quantum information}

\newacronym{irrep}{{\rm irrep}}{irreducible representation}

\newacronym{ASQR}{asqr}{antisymmetric qubit regularization}

%% file: header-4-title+authors.tex
\title{Symmetric Mass Generation as a Multicritical Point with Enhanced Symmetry}
\author{Sandip Maiti\,\orcidlink{0000-0002-5248-5316}}
\email{sandipmaiti73@gmail.com}
\affiliation{Key Laboratory of Quark and Lepton Physics (MOE) and Institute of Particle Physics, Central China Normal University, Wuhan 430079, China}
\affiliation{Saha Institute of Nuclear Physics, HBNI, 1/AF Bidhannagar, Kolkata 700064, India}
\author{Debasish Banerjee\,\orcidlink{0000-0003-0244-4337}}
\email{D.Banerjee@soton.ac.uk}
\affiliation{Saha Institute of Nuclear Physics, HBNI, 1/AF Bidhannagar, Kolkata 700064, India}
\affiliation{School of Physics and Astronomy, University of Southampton, University Road, SO17 1BJ, UK}
\author{Shailesh Chandrasekharan\,\orcidlink{0000-0002-3711-4998}}
\email{sch27@duke.edu}
\affiliation{Department of Physics, Box 90305, Duke University, Durham, North Carolina 27708, USA}
\author{Marina K.~Marinkovic\,\orcidlink{0000-0002-9883-7866}}
\email{marinama@ethz.ch}
\affiliation{Institut f\"ur Theoretische Physik, Wolfgang-Pauli-Stra{\ss}e 27, ETH Z\"urich, 8093 Z\"urich, Switzerland}

%% file: header-5-abstract.tex
\begin{abstract}
We explore the phase diagram of a lattice fermion model that exhibits three distinct phases: a massless fermion phase, a massive fermion phase with spontaneous symmetry breaking induced by a fermion bilinear condensate, and a massive fermion phase with symmetric mass generation. Using the fermion-bag Monte Carlo method on large cubical lattices, we find evidence for traditional second-order critical points separating the first two and the latter two phases. Remarkably, these critical points appear to merge at a multicritical point with enhanced symmetry when the symmetry breaking parameter is tuned to zero, giving rise to the recently discovered direct second-order transition between the massless and symmetric massive fermion phases.
\end{abstract}

%% file: sec-1.tex
In continuum quantum field theories, relativistic fermions typically acquire masses through explicit fermion bilinear terms in the Lagrangian. When such terms are forbidden by symmetries, fermion masses may still be generated dynamically through the spontaneous breaking of those symmetries~\cite{Gross:1974jv}. Beginning in the late 1980s, however, it was discovered that massive fermionic phases can emerge in certain lattice models with strictly massless fermions (MFs) and \emph{without} any spontaneous symmetry breaking (SSB), provided the interactions are sufficiently strong~\cite{Hasenfratz:1989jr,Lee:1989mi,Bock:1990tv}. These symmetric massive fermion phases were long regarded as lattice artifacts of little relevance to continuum quantum field theory, since the resulting fermion masses typically reside at the lattice cutoff scale.

This perspective has evolved significantly in recent years. Evidence for continuous transitions between the massless fermion phase and the symmetric massive fermion phase has now been found in both purely fermionic theories~\cite{Slagle:2014vma,Ayyar:2014eua,Ayyar:2015lrd,Catterall:2015zua,Ayyar:2016lxq,He:2016sbs} and in models that include gauge interactions~\cite{Butt:2021koj,Butt:2024kxi,Hasenfratz:2025lti}. At the same time, new insights into anomaly cancellation mechanisms that permit symmetric massive fermion phases have emerged~\cite{PhysRevX.11.011063,Tong:2021phe,Wang:2022ucy}.

Taken together, these developments suggest the presence of a new fixed point of the renormalization-group (RG) flow diagram, enabling a nontraditional mass-generation mechanism—now commonly referred to as the symmetric mass generation (SMG)—in continuum quantum field theory, whereby fermions acquire masses without spontaneous symmetry breaking \cite{PhysRevX.8.011026}. A perturbative study of the RG flows near this new SMG-fixed point using an $\epsilon$-expansion technique has recently been proposed~\cite{Martin:2025hic}. The SMG mechanism has also renewed interest in compositeness-based approaches to formulating chiral gauge theories on the lattice~\cite{Eichten:1985ft,Wang:2018ugf,Xu:2021ztz,Zeng:2022grc,Golterman:2023zqf,Golterman:2025boq,Mouland:2025ilu}.

\begin{figure}[t]
\centering
\includegraphics[width=0.48\textwidth]{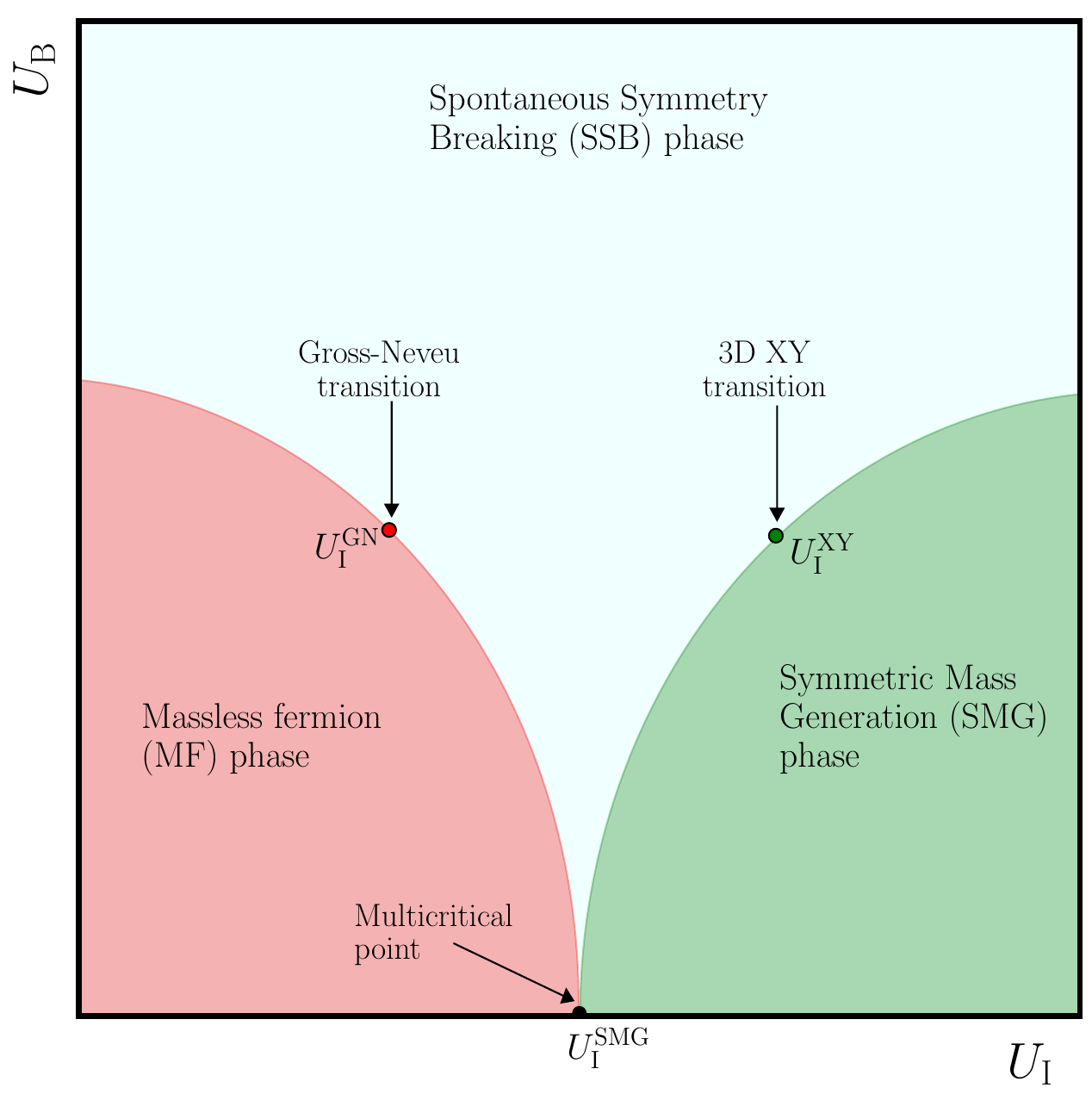}
\caption{Phase diagram of the two-flavor massless staggered fermion model in three space-time dimensions with two four-fermion couplings, as defined in \cref{eq:model-2}. The generic internal symmetry $\SU(2)\times \SU(2)\times U_\chi(1)$ is enhanced to $\SU(4)$ along the $U_B=0$ axis. The critical point on this axis, which appears as a direct second-order transition between the MF and SMG phases, is in fact a multicritical point where the Gross--Neveu and 3D XY critical lines merge. In the SSB phase the $U_\chi(1)$ symmetry is broken spontaneously.}
\label{fig:phase-diag}
\end{figure}

\begin{figure*}[t]
\centering
\includegraphics[width=0.45\textwidth]{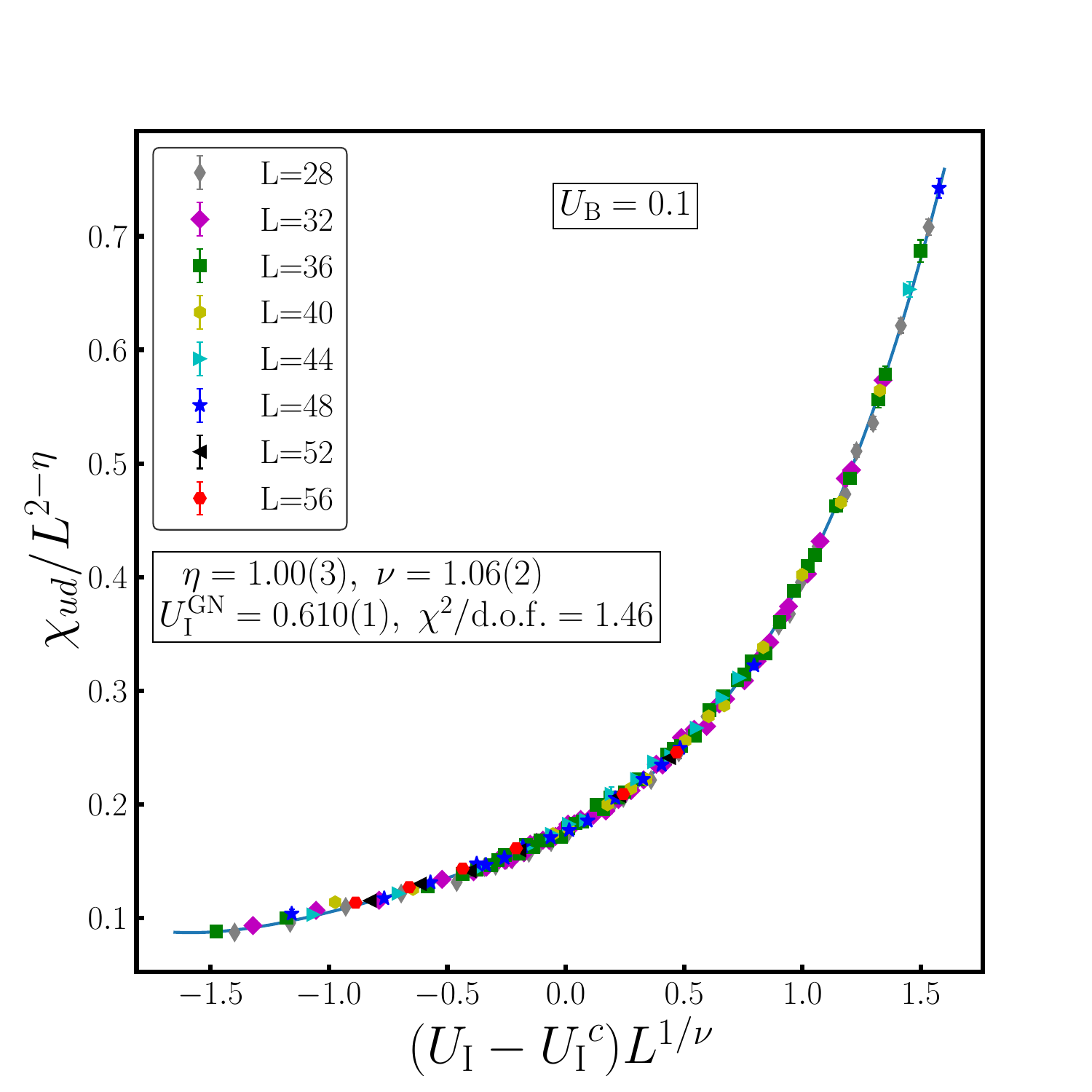}
\includegraphics[width=0.45\textwidth]{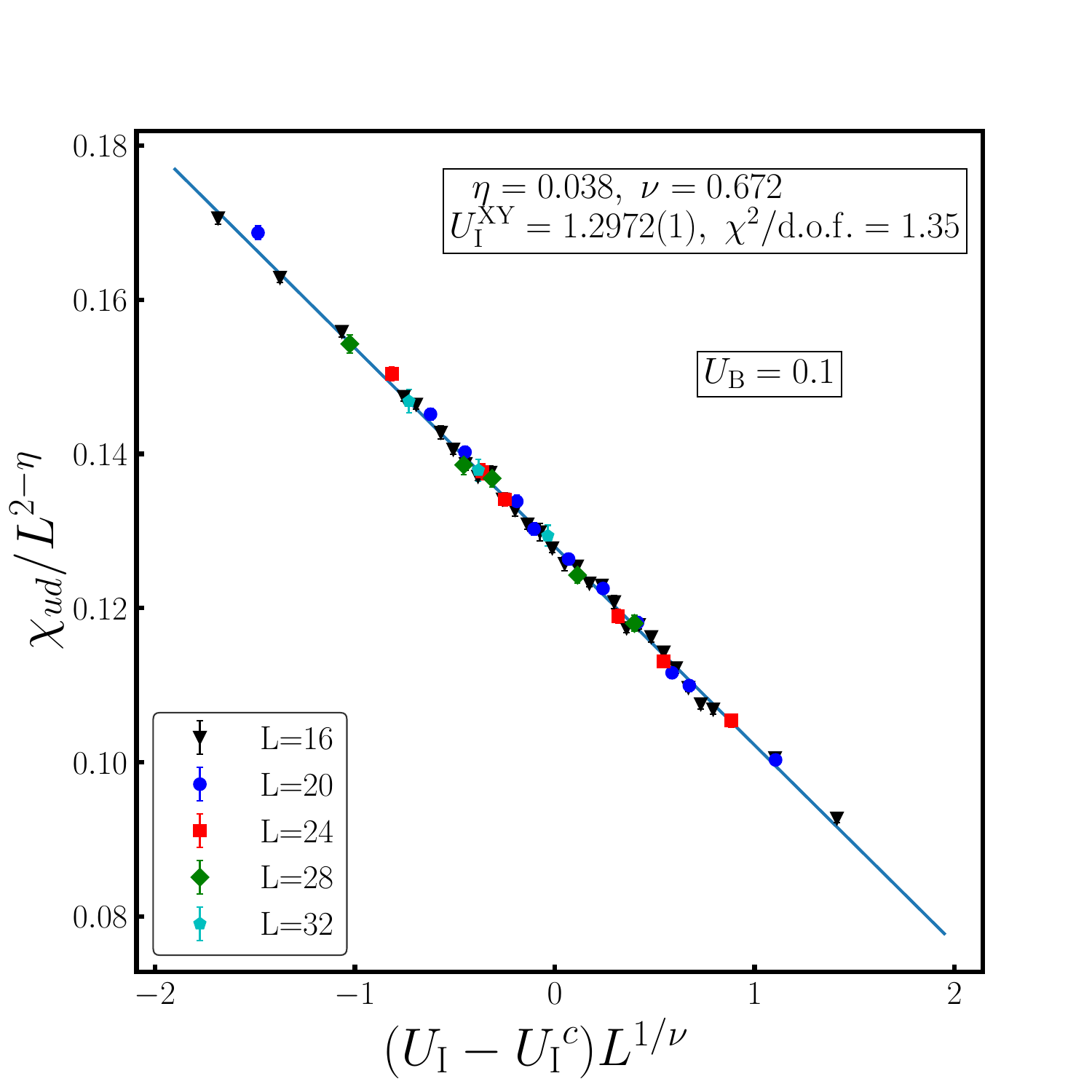}
\caption{Plots of the critical finite-size scaling function defined in \cref{eq:crit-scaling} at the Gross--Neveu transition (left panel) and the 3D-XY transition (right panel) for $U_B = 0.1$. The solid lines represent polynomial fits: a fourth-order polynomial for the Gross--Neveu case and a linear fit for the 3D-XY case. For the Gross--Neveu transition, the critical coupling and critical exponents are treated as fit parameters, whereas for the 3D-XY transition the critical exponents are fixed to their known values. The Monte Carlo data, shown with error bars, were obtained for various lattice sizes $L$ and a range of couplings $U_I$ in the vicinity of the respective critical points.}
\label{fig:FSS}
\end{figure*}

When a lattice model exhibits both a MF and a SMG phase, the existence of an intermediate phase—where fermions acquire masses through the conventional mechanism of SSB accompanied by a fermion bilinear condensate—cannot be excluded in two or more spatial dimensions. From this perspective, a direct second-order transition between the MF and SMG phases would ordinarily require fine tuning of at least two independent parameters, as originally argued in Ref.~\cite{Hasenfratz:1988vc}. It is therefore surprising that in a simple Euclidean lattice model with two flavors of massless staggered fermions, tuning a single local four-fermion coupling $U_I$ appeared to produce such a direct transition~\cite{Ayyar:2014eua,Ayyar:2015lrd}. 

The action of the model that was studied is given by
\begin{align}
S_1 
&= \sum_{\langle ij\rangle} \frac{\eta_{ij}}{2}\,\bigl(\ubar_i u_j - \ubar_j u_i + \dbar_i d_j - \dbar_j d_i
\bigr)
\notag \\
& \;-\; U_I \sum_{i} \bigl( \ubar_i u_i \,\dbar_i d_i \bigr),
\label{eq:model-1}
\end{align}
where $i$ and $j$ label sites of a three-dimensional cubic lattice, and $\langle ij\rangle$ denotes nearest-neighbor pairs. The staggered fermion fields $\ubar_i$, $u_i$, $\dbar_i$, and $d_i$ are Grassmann-valued, while the phase factors $\eta_{ij}$ implement a $\pi$-flux through every plaquette of the lattice. We impose anti-periodic boundary conditions in all three directions. In the continuum limit, these staggered fermions describe four massless flavors of four-component Dirac fermions~\cite{Sharatchandra:1981si}.

\begin{figure*}[t]
\centering
\includegraphics[trim=150 0 110 50, clip, width=0.96\textwidth]{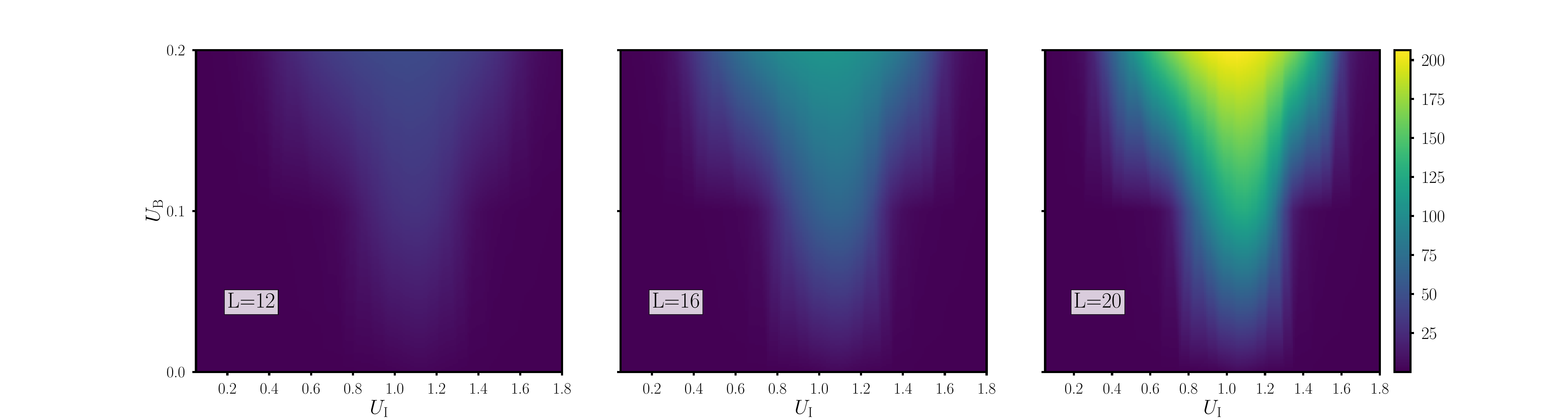}
\caption{Evidence for the phase diagram shown in \cref{fig:phase-diag}. Shown are heat maps of the susceptibility $\chi_{ud}$ at three lattice sizes, illustrating the structure of the phase diagram. Enhanced susceptibilities indicative of an SSB phase are visible for $U_B$ values as small as $U_B = 0.01$ at $U_I = 1.0$ (see \cite{Maiti:2026log}).}
\label{fig:heat-map}
\end{figure*}

An important feature of the model in \cref{eq:model-1} is its $\SU(4)$ internal symmetry:
\begin{align}
\begin{pmatrix} u_i \\[4pt] \ubar_i \\[4pt] d_i \\[4pt] \dbar_i \end{pmatrix}
&\rightarrow
V
\begin{pmatrix} u_i \\[4pt] \ubar_i \\[4pt] d_i \\[4pt] \dbar_i \end{pmatrix},
&
\begin{pmatrix} \ubar_j \\[4pt] u_j \\[4pt] \dbar_j \\[4pt] d_j \end{pmatrix}
&\rightarrow
V^{*}
\begin{pmatrix} \ubar_j \\[4pt] u_j \\[4pt] \dbar_j \\[4pt] d_j \end{pmatrix},
\label{eq:su4sym}
\end{align}
where $V$ is a $4\times4$ matrix in $\SU(4)$. Throughout this work we assign each lattice site a parity, with $i$ denoting an even site and $j$ an odd site. As we show below, this enlarged $\SU(4)$ symmetry plays a crucial role in eliminating the need to fine-tune two independent couplings in order to reach the direct second-order transition.

To expose the role of the $\SU(4)$ symmetry, we extend \cref{eq:model-1} by introducing an additional interaction $U_B$ that explicitly breaks it. The resulting action is
\begin{align}
S_2 &= S_1 \;-\; U_B \sum_{\langle ij\rangle}
\bigl( \ubar_i u_i\, \ubar_j u_j
      + \dbar_i d_i\, \dbar_j d_j \bigr).
\label{eq:model-2}
\end{align}
For $U_B\neq0$, the $\SU(4)$ symmetry is reduced to
$\SU(2)\times\SU(2)\times U_\chi(1)$. One $\SU(2)$ factor acts on the $(u,\ubar)$ doublet according to
\begin{align}
\begin{pmatrix} u_i \\[4pt] \ubar_i \end{pmatrix}
&\rightarrow
G_u
\begin{pmatrix} u_i \\[4pt] \ubar_i \end{pmatrix},
&
\begin{pmatrix} \ubar_j \\[4pt] u_j \end{pmatrix}
&\rightarrow
G_u^{*}
\begin{pmatrix} \ubar_j \\[4pt] u_j \end{pmatrix},
\label{eq:su2sym}
\end{align}
where $G_u$ is a $2\times2$ matrix in $\SU(2)$. An independent matrix $G_d\in\SU(2)$ acts analogously on the $(d,\dbar)$ doublet. The chiral symmetry $U_\chi(1)$ acts in the same way, with $G_u=e^{i\theta}$ and $G_d=e^{-i\theta}$. As before, $i$ and $j$ denote even and odd lattice sites, respectively.

Using an extension of the fermion bag algorithm~\cite{Chandrasekharan:2009wc}, we map out the phase diagram of \cref{eq:model-2} in the $(U_I,U_B)$ plane. We find all three expected phases: the MF phase at small $U_I$ and $U_B$, the SMG phase at large $U_I$ and small $U_B$, and an intermediate SSB phase separating them, as shown in \cref{fig:phase-diag}. The SSB phase is characterized by spontaneous breaking of the $U_\chi(1)$ symmetry.

The resulting phase diagram shows that the direct second-order transition at $U_B=0$, reported in earlier studies, is a consequence of the enlarged $\SU(4)$ symmetry. Specifically, the point at $U_B=0$ is naturally interpreted as a multicritical point at which the critical line separating the MF and SSB phases meets the critical line separating the SSB and SMG phases. The former transition is consistent with the three-dimensional Gross--Neveu $U(1)$ universality class with $N_f=4$ flavors of four-component Dirac fermions, while the latter belongs to the three-dimensional bosonic XY universality class.

The calculations reported here are performed on some of the largest lattices studied for this class of models. For the Gross--Neveu transition we simulate cubic lattices with linear size up to $L=56$, while for the bosonic XY transition---which is particularly challenging to access in a microscopic fermionic model---we reach $L=32$. Details of the extended fermion bag Monte Carlo algorithm, together with the numerical results and finite-size scaling analyses, are presented in the companion paper~\cite{Maiti:2026log}.

%% file: sec-3.tex
One of the observables we compute using our Monte Carlo method to study the phase structure is the susceptibility involving the $u$ and $d$ fields,
\begin{align}
\chi_{ud} \;=\; \frac{1}{2V}\sum_{i,j} \big\langle \ubar_i u_i \;\dbar_j d_j \big\rangle,
\label{eq:obs-chiud}
\end{align}
where $\langle \cdot \rangle$ denotes the path-integral expectation value using the action in \cref{eq:model-2}. By replacing the $d$ fields with $u$ fields in \cref{eq:obs-chiud}, or vice versa, we obtain two additional susceptibilities, $\chi_{uu}$ and $\chi_{dd}$. It is easy to verify that $\chi_{uu} = \chi_{dd}$. We also find that $\chi_{ud}$ and $\chi_{uu}$ are approximately equal on large lattices except for small $U_I$. Further details regarding the behavior of these susceptibilities are provided in Ref.~\cite{Maiti:2026log}.

Most of our detailed calculations related to the two transitions were performed at $U_B = 0.1$ for various values of $U_I$ near the Gross--Neveu critical point $U_I^{\rm GN}$ and the 3D--XY critical point $U_I^{\rm XY}$, as indicated in \cref{fig:phase-diag}. We discuss our results for the two transitions in detail below. Near both transitions we fit the susceptibility data to the critical scaling form
\begin{align}
\chi\,/ L^{2-\eta} = f\!\left[(U_I - U_I^c)\, L^{1/\nu}\right],
\label{eq:crit-scaling}
\end{align}
where $\chi$ denotes either $\chi_{ud}$ or $\chi_{uu}$, with each having its own universal scaling function $f(x)$, where $x = (U_I - U_I^c)\, L^{1/\nu}$.

At the Gross--Neveu transition we set $U_I^c = U_I^{\rm GN}$ and determine this value together with the corresponding exponents $\eta$ and $\nu$ by performing a combined fit of both $\chi_{ud}$ and $\chi_{uu}$ data to the scaling form \cref{eq:crit-scaling}, using a fourth-order polynomial expansion
\begin{align}
f(x) \;=\; 
f_0 + f_1 x + f_2 x^2 + f_3 x^3 + f_4 x^4,
\label{eq:univfn-GN}
\end{align}
for each susceptibility. Our results for $\chi_{ud}$ and the corresponding fit are shown in the left panel of \cref{fig:FSS}. 
Statistical errors in the Monte Carlo data have been computed using the jackknife method, and cross-checked using bootstrap as well as autocorrelation corrected errors for consistency. We obtain $U_I^{\rm GN}=0.610(1)$, $\nu=1.06(2)$, and $\eta=1.00(3)$. The polynomial coefficients entering the universal function in \cref{eq:univfn-GN} for $\chi_{ud}$ and $\chi_{uu}$ are listed in \cref{tab:fitparam-GN}.

\begin{table}[h]
\centering
\renewcommand{\arraystretch}{1.4}
\setlength{\tabcolsep}{4pt}
\begin{tabular}{c|c|c|c|c|c}
\TopRule
& $f_0$ & $f_1$ & $f_2$ & $f_3$ & $f_4$ \\
\MidRule
$\chi_{ud}$ & 0.18(2) & 0.11(2) & 0.06(1) & 0.04(1) & 0.014(4)
\\
$\chi_{uu}$ & 0.19(1) & 0.11(1) & 0.06(1) & 0.04(1) & 0.015(4) 
\\
\BotRule 
\end{tabular}
\caption{
The first five polynomial coefficients of the universal critical function in \cref{eq:univfn-GN} for both $\chi_{ud}$ and $\chi_{uu}$. The critical coupling and exponents were included as fit parameters and found to be $U_I^{\rm GN} = 0.610(1)$, $\nu = 1.06(2)$, and $\eta = 1.00(3)$. The data and corresponding fit are shown in the left panel of \cref{fig:FSS}.
\label{tab:fitparam-GN}
}
\end{table}

The critical exponents we obtain agree within errors with the four-loop $\epsilon$-expansion results~\cite{PhysRevD.96.096010}.
However, this comparison should be treated with caution, since the embedding of the symmetries of three-dimensional Euclidean staggered fermions into those of the continuum theory is non-trivial (see, for example, \cite{Gehring2015,Hands2021}).
It is also worth noting that these values lie close to the large-$N_f$ mean-field predictions $\nu = 1$ and $\eta = 1$~\cite{Hands:1992be}. This Gross--Neveu transition is additionally relevant in the context of twisted bilayer graphene; recent Monte Carlo studies on much smaller lattices report similar exponent values~\cite{Nature2025-Huang}. Other recent theoretical studies also find consistent numbers with small variations. Table I of ~\cite{PhysRevB.111.205129} lists the latest results from several studies.

The presence of the SMG phase at large values of $U_I$ implies a second phase transition, at which the spontaneous breaking of the $U(1)_\chi$ symmetry in the SSB phase is lost while the fermions remain massive and therefore decouple from the critical dynamics. This transition is thus expected to belong to the 3D--XY universality class. Denoting the corresponding critical coupling by $U_I^c = U_I^{\rm XY}$, we determine its value by performing a combined fit of the $\chi_{ud}$ and $\chi_{uu}$ data to \cref{eq:crit-scaling}. In this case, a linear approximation to the universal scaling function,
\begin{align}
f(x) = f_0 + f_1 x,
\label{eq:univfn-XY}
\end{align}
was found to be sufficient. Including higher-order polynomial terms does not improve the fit quality within statistical uncertainties. In the fit, we fix the critical exponents to their known 3D--XY values, $\nu = 0.6715$ and $\eta = 0.0380$~\cite{PhysRevB.63.214503}. Our results for $\chi_{ud}$, together with the corresponding fit, are shown in the right panel of \cref{fig:FSS}, and the fitted parameter values are listed in \cref{tab:fitparam-XY}.

\begin{table}[t]
\centering
\renewcommand{\arraystretch}{1.4}
\setlength{\tabcolsep}{4pt}
\begin{tabular}{c|c|c}
\TopRule
& $f_0$ & $f_1$ \\
\MidRule
$\chi_{ud}$ & 0.1280(2) & -0.0258(1) 
\\
$\chi_{uu}$ & 0.1287(2) & -0.0257(1)
\\
\BotRule 
\end{tabular}
\caption{
The first two coefficients of the polynomial expansion of the universal critical function in \cref{eq:univfn-XY} for both $\chi_{ud}$ and $\chi_{uu}$. The exponents $\nu=0.6715$ and $\eta=0.0380$ were fixed to their known 3D values, but the critical coupling $U_I^{XY} = 1.2972(1)$ was obtained in the fit. The data and corresponding fit are shown in the right panel of \cref{fig:FSS}.
\label{tab:fitparam-XY}
}
\end{table}

\begin{figure}[h]
\centering
\includegraphics[width=0.45\textwidth]{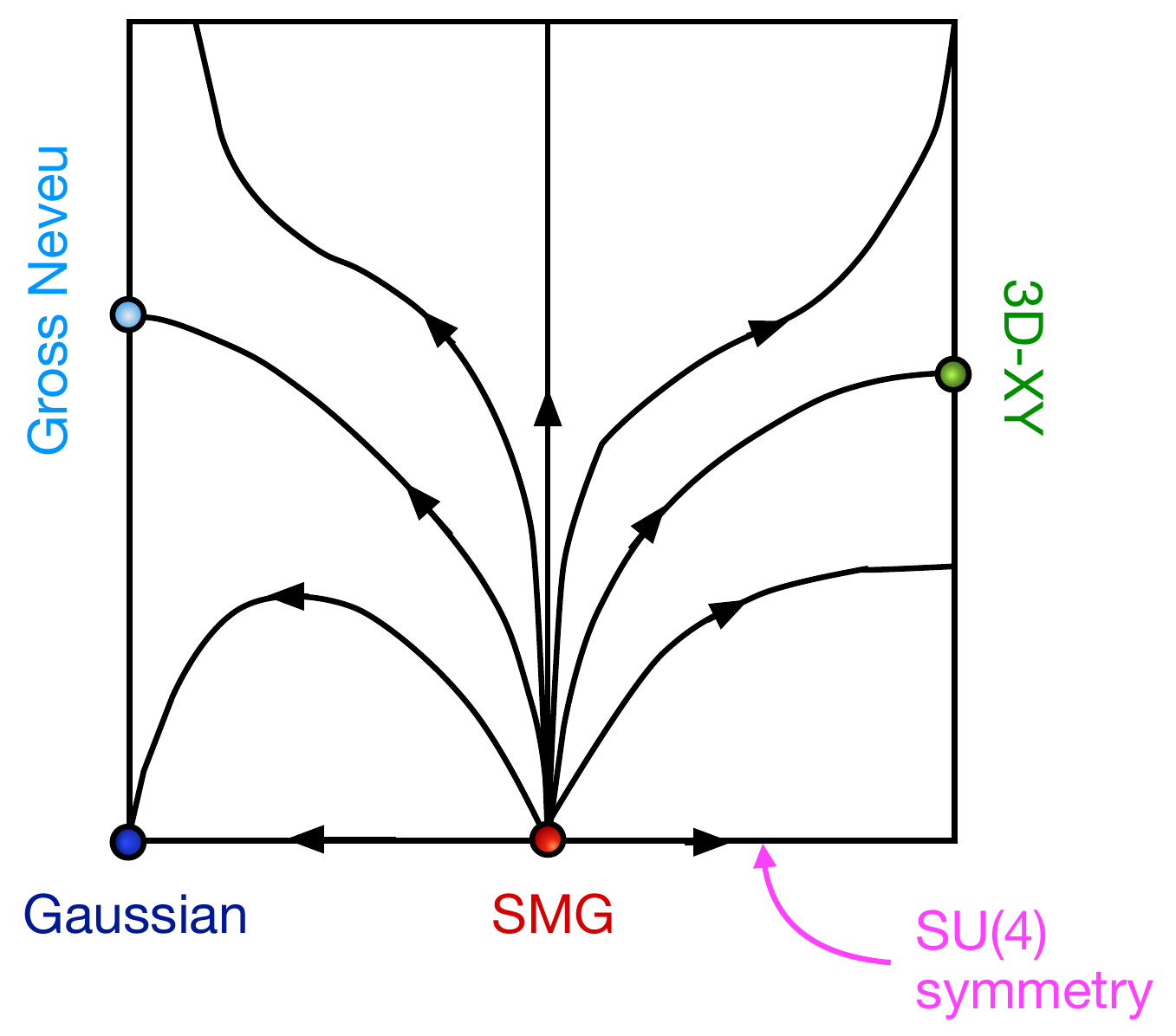}
\caption{Conjectured RG flows for the three-dimensional fermionic Gross--Neveu--Yukawa models near the SMG fixed point. Our results indicate four fixed points: the Gaussian fixed point (massless fermions, massive bosons); a Gross--Neveu fixed point (interacting massless fermions and bosons, one relevant direction); an XY fixed point (interacting massless bosons, one relevant direction); and the new SMG fixed point, which has two relevant directions but reduces to one when an additional symmetry is imposed.}
\label{fig:RGflows}
\end{figure}

%% file: sec-4.tex
To establish that the phase diagram of \cref{eq:model-2} is indeed given by \cref{fig:phase-diag}, we computed the susceptibilities for various values of $U_B$ and $U_I$ at $L = 12$, $L = 16$, and $L = 20$. The resulting heat maps are shown in \cref{fig:heat-map}. Darker colors correspond to smaller susceptibility values, while lighter colors indicate larger values, with the color scale displayed alongside each plot. These results are consistent with the phase structure summarized in \cref{fig:phase-diag}.

Based on the results presented here, we find strong support for the interpretation that the 3D Gross--Neveu and 3D--XY critical lines studied in this work merge at a multicritical point, identified as the SMG critical point discovered in earlier studies~\cite{Ayyar:2014eua,Ayyar:2015lrd}. This scenario implies the existence of at least four distinct fixed points in the RG flow of the associated Gross--Neveu--Yukawa theories with four flavors of four-component massless Dirac fermions, one of which is the SMG fixed point. Our conjectured RG flows are illustrated in \cref{fig:RGflows}, and closely resemble those proposed long ago in \cite{Hasenfratz:1988vc}. The enhanced $\SU(4)$ symmetry at the SMG fixed point plays a crucial role, as it allows the multicritical point to be accessed by tuning a single coupling rather than two, thereby facilitating its identification in earlier lattice studies in three space-time dimensions~\cite{Ayyar:2014eua,Ayyar:2015lrd}.

Recent perturbative analyses of RG flows near the SMG fixed point have focused on $\SU(4)$-invariant theories \cite{Martin:2025hic}. It should, however, be straightforward to extend these studies by including symmetry-breaking terms that reduce the symmetry to $\SU(2)\times \SU(2)\times U_\chi(1)$ and allow for spontaneous breaking of the $U_\chi(1)$ symmetry. Such an extension would enable a direct test of the RG-flow structure proposed in \cref{fig:RGflows}.

It would be interesting to investigate whether the multicriticality identified in our work also emerges in the Hamiltonian formulation of the SMG transition studied in \cite{He:2016sbs}, and in four space-time dimensions. In the latter case, since \cref{eq:model-2} can also be extended to four space-time dimensions, the intermediate SSB phase is known to persist even at $U_B = 0$~\cite{Ayyar:2016lxq}. 
There are preliminary indications that an SMG fixed point may also exist in four space-time dimensions when the model is extended to include Yukawa interactions, and two parameters are fine tuned~\cite{Butt:2018nkn}. 
Clarifying this issue and testing the persistence of any enhanced multicritical structure beyond three dimensions remains an important direction for future work.

%% file: acknowledgments.tex
\section*{Acknowledgments}

We are grateful to Subhro Bhattacharjee, Simon Catterall, Asit De, Simon Hands, and Cenke Xu for insightful discussions. S.C. is supported in part by the U.S. Department of Energy, Office of Science, Nuclear Physics program under Award No.~DE-FG02-05ER41368. This research was supported in part by the National Science Foundation under Grant No. NSF PHY-1748958. 
D.B. would like to thank STFC (UK) consolidated Grant No. ST/X000583/1 and continued support from the Alexander von Humboldt Foundation (Germany) in the context of the research fellowship for experienced researcher. M.K.M. is grateful for the hospitality of Perimeter Institute where part of this work was carried out. Research at Perimeter Institute is supported in part by the Government of Canada through the Department of Innovation, Science and Economic Development and by the Province of Ontario through the Ministry of Colleges and Universities. This research was also supported in part by the Simons Foundation through the Simons Foundation Emmy Noether Fellows Program at Perimeter Institute. S.M. would like to thank the Institute for Theoretical Physics at ETHZ for hospitality during his visit, where part of this work was carried out. S.M. would like to acknowledge support from the National Natural Science Foundation of China under Grants No.~12325508, No.~12293064, and No.~12293060. D.B., S.M., and M.K.M. also acknowledge the use of computing clusters at SINP, Kolkata, and access to the Piz Daint and Eiger supercomputers at the Swiss National Supercomputing Centre, Switzerland, under ETHZ’s allocation with project IDs \texttt{c21} and \texttt{eth8}. We acknowledge the use of AI assistance, specifically ChatGPT~\hbox{\cite{openai2025chatgpt}}, in refining the language and clarity of this manuscript.

%% file: paper.bbl
\begin{thebibliography}{38}%
\makeatletter
\providecommand \@ifxundefined [1]{%
 \@ifx{#1\undefined}
}%
\providecommand \@ifnum [1]{%
 \ifnum #1\expandafter \@firstoftwo
 \else \expandafter \@secondoftwo
 \fi
}%
\providecommand \@ifx [1]{%
 \ifx #1\expandafter \@firstoftwo
 \else \expandafter \@secondoftwo
 \fi
}%
\providecommand \natexlab [1]{#1}%
\providecommand \enquote  [1]{``#1''}%
\providecommand \bibnamefont  [1]{#1}%
\providecommand \bibfnamefont [1]{#1}%
\providecommand \citenamefont [1]{#1}%
\providecommand \href@noop [0]{\@secondoftwo}%
\providecommand \href [0]{\begingroup \@sanitize@url \@href}%
\providecommand \@href[1]{\@@startlink{#1}\@@href}%
\providecommand \@@href[1]{\endgroup#1\@@endlink}%
\providecommand \@sanitize@url [0]{\catcode `\\12\catcode `\$12\catcode `\&12\catcode `\#12\catcode `\^12\catcode `\_12\catcode `\%12\relax}%
\providecommand \@@startlink[1]{}%
\providecommand \@@endlink[0]{}%
\providecommand \url  [0]{\begingroup\@sanitize@url \@url }%
\providecommand \@url [1]{\endgroup\@href {#1}{\urlprefix }}%
\providecommand \urlprefix  [0]{URL }%
\providecommand \Eprint [0]{\href }%
\providecommand \doibase [0]{https://doi.org/}%
\providecommand \selectlanguage [0]{\@gobble}%
\providecommand \bibinfo  [0]{\@secondoftwo}%
\providecommand \bibfield  [0]{\@secondoftwo}%
\providecommand \translation [1]{[#1]}%
\providecommand \BibitemOpen [0]{}%
\providecommand \bibitemStop [0]{}%
\providecommand \bibitemNoStop [0]{.\EOS\space}%
\providecommand \EOS [0]{\spacefactor3000\relax}%
\providecommand \BibitemShut  [1]{\csname bibitem#1\endcsname}%
\let\auto@bib@innerbib\@empty
\bibitem [{\citenamefont {Gross}\ and\ \citenamefont {Neveu}(1974)}]{Gross:1974jv}%
  \BibitemOpen
  \bibfield  {author} {\bibinfo {author} {\bibfnamefont {D.~J.}\ \bibnamefont {Gross}}\ and\ \bibinfo {author} {\bibfnamefont {A.}~\bibnamefont {Neveu}},\ }\bibfield  {title} {\bibinfo {title} {{Dynamical Symmetry Breaking in Asymptotically Free Field Theories}},\ }\href {https://doi.org/10.1103/PhysRevD.10.3235} {\bibfield  {journal} {\bibinfo  {journal} {Phys. Rev. D}\ }\textbf {\bibinfo {volume} {10}},\ \bibinfo {pages} {3235} (\bibinfo {year} {1974})}\BibitemShut {NoStop}%
\bibitem [{\citenamefont {Hasenfratz}\ \emph {et~al.}(1990)\citenamefont {Hasenfratz}, \citenamefont {Liu},\ and\ \citenamefont {Neuhaus}}]{Hasenfratz:1989jr}%
  \BibitemOpen
  \bibfield  {author} {\bibinfo {author} {\bibfnamefont {A.}~\bibnamefont {Hasenfratz}}, \bibinfo {author} {\bibfnamefont {W.-q.}\ \bibnamefont {Liu}},\ and\ \bibinfo {author} {\bibfnamefont {T.}~\bibnamefont {Neuhaus}},\ }\bibfield  {title} {\bibinfo {title} {{Phase Structure and Critical Points in a Scalar Fermion Model}},\ }\href {https://doi.org/10.1016/0370-2693(90)90994-H} {\bibfield  {journal} {\bibinfo  {journal} {Phys. Lett. B}\ }\textbf {\bibinfo {volume} {236}},\ \bibinfo {pages} {339} (\bibinfo {year} {1990})}\BibitemShut {NoStop}%
\bibitem [{\citenamefont {Lee}\ \emph {et~al.}(1990)\citenamefont {Lee}, \citenamefont {Shigemitsu},\ and\ \citenamefont {Shrock}}]{Lee:1989mi}%
  \BibitemOpen
  \bibfield  {author} {\bibinfo {author} {\bibfnamefont {I.-H.}\ \bibnamefont {Lee}}, \bibinfo {author} {\bibfnamefont {J.}~\bibnamefont {Shigemitsu}},\ and\ \bibinfo {author} {\bibfnamefont {R.~E.}\ \bibnamefont {Shrock}},\ }\bibfield  {title} {\bibinfo {title} {{Study of Different Lattice Formulations of a Yukawa Model With a Real Scalar Field}},\ }\href {https://doi.org/10.1016/0550-3213(90)90664-Y} {\bibfield  {journal} {\bibinfo  {journal} {Nucl. Phys. B}\ }\textbf {\bibinfo {volume} {334}},\ \bibinfo {pages} {265} (\bibinfo {year} {1990})}\BibitemShut {NoStop}%
\bibitem [{\citenamefont {Bock}\ \emph {et~al.}(1990)\citenamefont {Bock}, \citenamefont {De}, \citenamefont {Jansen}, \citenamefont {Jersak}, \citenamefont {Neuhaus},\ and\ \citenamefont {Smit}}]{Bock:1990tv}%
  \BibitemOpen
  \bibfield  {author} {\bibinfo {author} {\bibfnamefont {W.}~\bibnamefont {Bock}}, \bibinfo {author} {\bibfnamefont {A.~K.}\ \bibnamefont {De}}, \bibinfo {author} {\bibfnamefont {K.}~\bibnamefont {Jansen}}, \bibinfo {author} {\bibfnamefont {J.}~\bibnamefont {Jersak}}, \bibinfo {author} {\bibfnamefont {T.}~\bibnamefont {Neuhaus}},\ and\ \bibinfo {author} {\bibfnamefont {J.}~\bibnamefont {Smit}},\ }\bibfield  {title} {\bibinfo {title} {{Phase Diagram of a Lattice SU(2) X SU(2) Scalar Fermion Model With Naive and Wilson Fermions}},\ }\href {https://doi.org/10.1016/0550-3213(90)90689-B} {\bibfield  {journal} {\bibinfo  {journal} {Nucl. Phys. B}\ }\textbf {\bibinfo {volume} {344}},\ \bibinfo {pages} {207} (\bibinfo {year} {1990})}\BibitemShut {NoStop}%
\bibitem [{\citenamefont {Slagle}\ \emph {et~al.}(2015)\citenamefont {Slagle}, \citenamefont {You},\ and\ \citenamefont {Xu}}]{Slagle:2014vma}%
  \BibitemOpen
  \bibfield  {author} {\bibinfo {author} {\bibfnamefont {K.}~\bibnamefont {Slagle}}, \bibinfo {author} {\bibfnamefont {Y.-Z.}\ \bibnamefont {You}},\ and\ \bibinfo {author} {\bibfnamefont {C.}~\bibnamefont {Xu}},\ }\bibfield  {title} {\bibinfo {title} {{Exotic quantum phase transitions of strongly interacting topological insulators}},\ }\href {https://doi.org/10.1103/PhysRevB.91.115121} {\bibfield  {journal} {\bibinfo  {journal} {Phys. Rev. B}\ }\textbf {\bibinfo {volume} {91}},\ \bibinfo {pages} {115121} (\bibinfo {year} {2015})},\ \Eprint {https://arxiv.org/abs/1409.7401} {arXiv:1409.7401 [cond-mat.str-el]} \BibitemShut {NoStop}%
\bibitem [{\citenamefont {Ayyar}\ and\ \citenamefont {Chandrasekharan}(2015)}]{Ayyar:2014eua}%
  \BibitemOpen
  \bibfield  {author} {\bibinfo {author} {\bibfnamefont {V.}~\bibnamefont {Ayyar}}\ and\ \bibinfo {author} {\bibfnamefont {S.}~\bibnamefont {Chandrasekharan}},\ }\bibfield  {title} {\bibinfo {title} {{Massive fermions without fermion bilinear condensates}},\ }\href {https://doi.org/10.1103/PhysRevD.91.065035} {\bibfield  {journal} {\bibinfo  {journal} {Phys. Rev. D}\ }\textbf {\bibinfo {volume} {91}},\ \bibinfo {pages} {065035} (\bibinfo {year} {2015})},\ \Eprint {https://arxiv.org/abs/1410.6474} {arXiv:1410.6474 [hep-lat]} \BibitemShut {NoStop}%
\bibitem [{\citenamefont {Ayyar}\ and\ \citenamefont {Chandrasekharan}(2016{\natexlab{a}})}]{Ayyar:2015lrd}%
  \BibitemOpen
  \bibfield  {author} {\bibinfo {author} {\bibfnamefont {V.}~\bibnamefont {Ayyar}}\ and\ \bibinfo {author} {\bibfnamefont {S.}~\bibnamefont {Chandrasekharan}},\ }\bibfield  {title} {\bibinfo {title} {{Origin of fermion masses without spontaneous symmetry breaking}},\ }\href {https://doi.org/10.1103/PhysRevD.93.081701} {\bibfield  {journal} {\bibinfo  {journal} {Phys. Rev. D}\ }\textbf {\bibinfo {volume} {93}},\ \bibinfo {pages} {081701} (\bibinfo {year} {2016}{\natexlab{a}})},\ \Eprint {https://arxiv.org/abs/1511.09071} {arXiv:1511.09071 [hep-lat]} \BibitemShut {NoStop}%
\bibitem [{\citenamefont {Catterall}(2016)}]{Catterall:2015zua}%
  \BibitemOpen
  \bibfield  {author} {\bibinfo {author} {\bibfnamefont {S.}~\bibnamefont {Catterall}},\ }\bibfield  {title} {\bibinfo {title} {{Fermion mass without symmetry breaking}},\ }\href {https://doi.org/10.1007/JHEP01(2016)121} {\bibfield  {journal} {\bibinfo  {journal} {JHEP}\ }\textbf {\bibinfo {volume} {01}},\ \bibinfo {pages} {121}},\ \Eprint {https://arxiv.org/abs/1510.04153} {arXiv:1510.04153 [hep-lat]} \BibitemShut {NoStop}%
\bibitem [{\citenamefont {Ayyar}\ and\ \citenamefont {Chandrasekharan}(2016{\natexlab{b}})}]{Ayyar:2016lxq}%
  \BibitemOpen
  \bibfield  {author} {\bibinfo {author} {\bibfnamefont {V.}~\bibnamefont {Ayyar}}\ and\ \bibinfo {author} {\bibfnamefont {S.}~\bibnamefont {Chandrasekharan}},\ }\bibfield  {title} {\bibinfo {title} {{Fermion masses through four-fermion condensates}},\ }\href {https://doi.org/10.1007/JHEP10(2016)058} {\bibfield  {journal} {\bibinfo  {journal} {JHEP}\ }\textbf {\bibinfo {volume} {10}},\ \bibinfo {pages} {058}},\ \Eprint {https://arxiv.org/abs/1606.06312} {arXiv:1606.06312 [hep-lat]} \BibitemShut {NoStop}%
\bibitem [{\citenamefont {He}\ \emph {et~al.}(2016)\citenamefont {He}, \citenamefont {Wu}, \citenamefont {You}, \citenamefont {Xu}, \citenamefont {Meng},\ and\ \citenamefont {Lu}}]{He:2016sbs}%
  \BibitemOpen
  \bibfield  {author} {\bibinfo {author} {\bibfnamefont {Y.-Y.}\ \bibnamefont {He}}, \bibinfo {author} {\bibfnamefont {H.-Q.}\ \bibnamefont {Wu}}, \bibinfo {author} {\bibfnamefont {Y.-Z.}\ \bibnamefont {You}}, \bibinfo {author} {\bibfnamefont {C.}~\bibnamefont {Xu}}, \bibinfo {author} {\bibfnamefont {Z.~Y.}\ \bibnamefont {Meng}},\ and\ \bibinfo {author} {\bibfnamefont {Z.-Y.}\ \bibnamefont {Lu}},\ }\bibfield  {title} {\bibinfo {title} {{Quantum critical point of Dirac fermion mass generation without spontaneous symmetry breaking}},\ }\href {https://doi.org/10.1103/PhysRevB.94.241111} {\bibfield  {journal} {\bibinfo  {journal} {Phys. Rev. B}\ }\textbf {\bibinfo {volume} {94}},\ \bibinfo {pages} {241111} (\bibinfo {year} {2016})},\ \Eprint {https://arxiv.org/abs/1603.08376} {arXiv:1603.08376 [cond-mat.str-el]} \BibitemShut {NoStop}%
\bibitem [{\citenamefont {Butt}\ \emph {et~al.}(2021)\citenamefont {Butt}, \citenamefont {Catterall},\ and\ \citenamefont {Toga}}]{Butt:2021koj}%
  \BibitemOpen
  \bibfield  {author} {\bibinfo {author} {\bibfnamefont {N.}~\bibnamefont {Butt}}, \bibinfo {author} {\bibfnamefont {S.}~\bibnamefont {Catterall}},\ and\ \bibinfo {author} {\bibfnamefont {G.~C.}\ \bibnamefont {Toga}},\ }\bibfield  {title} {\bibinfo {title} {{Symmetric Mass Generation in Lattice Gauge Theory}},\ }\href {https://doi.org/10.3390/sym13122276} {\bibfield  {journal} {\bibinfo  {journal} {Symmetry}\ }\textbf {\bibinfo {volume} {13}},\ \bibinfo {pages} {2276} (\bibinfo {year} {2021})},\ \Eprint {https://arxiv.org/abs/2111.01001} {arXiv:2111.01001 [hep-lat]} \BibitemShut {NoStop}%
\bibitem [{\citenamefont {Butt}\ \emph {et~al.}(2025)\citenamefont {Butt}, \citenamefont {Catterall},\ and\ \citenamefont {Hasenfratz}}]{Butt:2024kxi}%
  \BibitemOpen
  \bibfield  {author} {\bibinfo {author} {\bibfnamefont {N.}~\bibnamefont {Butt}}, \bibinfo {author} {\bibfnamefont {S.}~\bibnamefont {Catterall}},\ and\ \bibinfo {author} {\bibfnamefont {A.}~\bibnamefont {Hasenfratz}},\ }\bibfield  {title} {\bibinfo {title} {{Symmetric Mass Generation with Four SU(2) Doublet Fermions}},\ }\href {https://doi.org/10.1103/PhysRevLett.134.031602} {\bibfield  {journal} {\bibinfo  {journal} {Phys. Rev. Lett.}\ }\textbf {\bibinfo {volume} {134}},\ \bibinfo {pages} {031602} (\bibinfo {year} {2025})},\ \Eprint {https://arxiv.org/abs/2409.02062} {arXiv:2409.02062 [hep-lat]} \BibitemShut {NoStop}%
\bibitem [{\citenamefont {Hasenfratz}\ and\ \citenamefont {Witzel}(2025)}]{Hasenfratz:2025lti}%
  \BibitemOpen
  \bibfield  {author} {\bibinfo {author} {\bibfnamefont {A.}~\bibnamefont {Hasenfratz}}\ and\ \bibinfo {author} {\bibfnamefont {O.}~\bibnamefont {Witzel}} (\bibinfo {collaboration} {Lattice Strong Dynamics}),\ }\bibfield  {title} {\bibinfo {title} {{Symmetric Mass Generation}},\ }in\ \href@noop {} {\emph {\bibinfo {booktitle} {{2025 European Physical Society Conference on High Energy Physics}}}}\ (\bibinfo {year} {2025})\ \Eprint {https://arxiv.org/abs/2511.22678} {arXiv:2511.22678 [hep-lat]} \BibitemShut {NoStop}%
\bibitem [{\citenamefont {Razamat}\ and\ \citenamefont {Tong}(2021)}]{PhysRevX.11.011063}%
  \BibitemOpen
  \bibfield  {author} {\bibinfo {author} {\bibfnamefont {S.~S.}\ \bibnamefont {Razamat}}\ and\ \bibinfo {author} {\bibfnamefont {D.}~\bibnamefont {Tong}},\ }\bibfield  {title} {\bibinfo {title} {Gapped chiral fermions},\ }\href {https://doi.org/10.1103/PhysRevX.11.011063} {\bibfield  {journal} {\bibinfo  {journal} {Phys. Rev. X}\ }\textbf {\bibinfo {volume} {11}},\ \bibinfo {pages} {011063} (\bibinfo {year} {2021})}\BibitemShut {NoStop}%
\bibitem [{\citenamefont {Tong}(2022)}]{Tong:2021phe}%
  \BibitemOpen
  \bibfield  {author} {\bibinfo {author} {\bibfnamefont {D.}~\bibnamefont {Tong}},\ }\bibfield  {title} {\bibinfo {title} {{Comments on symmetric mass generation in 2d and 4d}},\ }\href {https://doi.org/10.1007/JHEP07(2022)001} {\bibfield  {journal} {\bibinfo  {journal} {JHEP}\ }\textbf {\bibinfo {volume} {07}},\ \bibinfo {pages} {001}},\ \Eprint {https://arxiv.org/abs/2104.03997} {arXiv:2104.03997 [hep-th]} \BibitemShut {NoStop}%
\bibitem [{\citenamefont {Wang}\ and\ \citenamefont {You}(2022)}]{Wang:2022ucy}%
  \BibitemOpen
  \bibfield  {author} {\bibinfo {author} {\bibfnamefont {J.}~\bibnamefont {Wang}}\ and\ \bibinfo {author} {\bibfnamefont {Y.-Z.}\ \bibnamefont {You}},\ }\bibfield  {title} {\bibinfo {title} {{Symmetric Mass Generation}},\ }\href {https://doi.org/10.3390/sym14071475} {\bibfield  {journal} {\bibinfo  {journal} {Symmetry}\ }\textbf {\bibinfo {volume} {14}},\ \bibinfo {pages} {1475} (\bibinfo {year} {2022})},\ \Eprint {https://arxiv.org/abs/2204.14271} {arXiv:2204.14271 [cond-mat.str-el]} \BibitemShut {NoStop}%
\bibitem [{\citenamefont {You}\ \emph {et~al.}(2018)\citenamefont {You}, \citenamefont {He}, \citenamefont {Xu},\ and\ \citenamefont {Vishwanath}}]{PhysRevX.8.011026}%
  \BibitemOpen
  \bibfield  {author} {\bibinfo {author} {\bibfnamefont {Y.-Z.}\ \bibnamefont {You}}, \bibinfo {author} {\bibfnamefont {Y.-C.}\ \bibnamefont {He}}, \bibinfo {author} {\bibfnamefont {C.}~\bibnamefont {Xu}},\ and\ \bibinfo {author} {\bibfnamefont {A.}~\bibnamefont {Vishwanath}},\ }\bibfield  {title} {\bibinfo {title} {Symmetric fermion mass generation as deconfined quantum criticality},\ }\href {https://doi.org/10.1103/PhysRevX.8.011026} {\bibfield  {journal} {\bibinfo  {journal} {Phys. Rev. X}\ }\textbf {\bibinfo {volume} {8}},\ \bibinfo {pages} {011026} (\bibinfo {year} {2018})}\BibitemShut {NoStop}%
\bibitem [{\citenamefont {Martin}\ and\ \citenamefont {Grover}(2025)}]{Martin:2025hic}%
  \BibitemOpen
  \bibfield  {author} {\bibinfo {author} {\bibfnamefont {S.}~\bibnamefont {Martin}}\ and\ \bibinfo {author} {\bibfnamefont {T.}~\bibnamefont {Grover}},\ }\bibfield  {title} {\bibinfo {title} {{A Perturbative Approach to Symmetric Mass Generation}},\ }\href@noop {} {\bibfield  {journal} {\bibinfo  {journal} {preprint}\ } (\bibinfo {year} {2025})},\ \Eprint {https://arxiv.org/abs/2507.23032} {arXiv:2507.23032 [cond-mat.str-el]} \BibitemShut {NoStop}%
\bibitem [{\citenamefont {Eichten}\ and\ \citenamefont {Preskill}(1986)}]{Eichten:1985ft}%
  \BibitemOpen
  \bibfield  {author} {\bibinfo {author} {\bibfnamefont {E.}~\bibnamefont {Eichten}}\ and\ \bibinfo {author} {\bibfnamefont {J.}~\bibnamefont {Preskill}},\ }\bibfield  {title} {\bibinfo {title} {{Chiral Gauge Theories on the Lattice}},\ }\href {https://doi.org/10.1016/0550-3213(86)90207-5} {\bibfield  {journal} {\bibinfo  {journal} {Nucl. Phys. B}\ }\textbf {\bibinfo {volume} {268}},\ \bibinfo {pages} {179} (\bibinfo {year} {1986})}\BibitemShut {NoStop}%
\bibitem [{\citenamefont {Wang}\ and\ \citenamefont {Wen}(2018)}]{Wang:2018ugf}%
  \BibitemOpen
  \bibfield  {author} {\bibinfo {author} {\bibfnamefont {J.}~\bibnamefont {Wang}}\ and\ \bibinfo {author} {\bibfnamefont {X.-G.}\ \bibnamefont {Wen}},\ }\bibfield  {title} {\bibinfo {title} {{A Solution to the 1+1D Gauged Chiral Fermion Problem}},\ }\href {https://doi.org/10.1103/PhysRevD.99.111501} {\bibfield  {journal} {\bibinfo  {journal} {Phys. Rev. D}\ }\textbf {\bibinfo {volume} {99}},\ \bibinfo {pages} {111501} (\bibinfo {year} {2018})},\ \Eprint {https://arxiv.org/abs/1807.05998} {arXiv:1807.05998 [hep-lat]} \BibitemShut {NoStop}%
\bibitem [{\citenamefont {Xu}\ and\ \citenamefont {Xu}(2021)}]{Xu:2021ztz}%
  \BibitemOpen
  \bibfield  {author} {\bibinfo {author} {\bibfnamefont {Y.}~\bibnamefont {Xu}}\ and\ \bibinfo {author} {\bibfnamefont {C.}~\bibnamefont {Xu}},\ }\bibfield  {title} {\bibinfo {title} {{Green's function Zero and Symmetric Mass Generation}},\ }\href@noop {} {\bibfield  {journal} {\bibinfo  {journal} {preprint}\ } (\bibinfo {year} {2021})},\ \Eprint {https://arxiv.org/abs/2103.15865} {arXiv:2103.15865 [cond-mat.str-el]} \BibitemShut {NoStop}%
\bibitem [{\citenamefont {Zeng}\ \emph {et~al.}(2022)\citenamefont {Zeng}, \citenamefont {Zhu}, \citenamefont {Wang},\ and\ \citenamefont {You}}]{Zeng:2022grc}%
  \BibitemOpen
  \bibfield  {author} {\bibinfo {author} {\bibfnamefont {M.}~\bibnamefont {Zeng}}, \bibinfo {author} {\bibfnamefont {Z.}~\bibnamefont {Zhu}}, \bibinfo {author} {\bibfnamefont {J.}~\bibnamefont {Wang}},\ and\ \bibinfo {author} {\bibfnamefont {Y.-Z.}\ \bibnamefont {You}},\ }\bibfield  {title} {\bibinfo {title} {{Symmetric Mass Generation in the 1+1 Dimensional Chiral Fermion 3-4-5-0 Model}},\ }\href {https://doi.org/10.1103/PhysRevLett.128.185301} {\bibfield  {journal} {\bibinfo  {journal} {Phys. Rev. Lett.}\ }\textbf {\bibinfo {volume} {128}},\ \bibinfo {pages} {185301} (\bibinfo {year} {2022})},\ \Eprint {https://arxiv.org/abs/2202.12355} {arXiv:2202.12355 [cond-mat.str-el]} \BibitemShut {NoStop}%
\bibitem [{\citenamefont {Golterman}\ and\ \citenamefont {Shamir}(2024)}]{Golterman:2023zqf}%
  \BibitemOpen
  \bibfield  {author} {\bibinfo {author} {\bibfnamefont {M.}~\bibnamefont {Golterman}}\ and\ \bibinfo {author} {\bibfnamefont {Y.}~\bibnamefont {Shamir}},\ }\bibfield  {title} {\bibinfo {title} {{Propagator Zeros and Lattice Chiral Gauge Theories}},\ }\href {https://doi.org/10.1103/PhysRevLett.132.081903} {\bibfield  {journal} {\bibinfo  {journal} {Phys. Rev. Lett.}\ }\textbf {\bibinfo {volume} {132}},\ \bibinfo {pages} {081903} (\bibinfo {year} {2024})},\ \Eprint {https://arxiv.org/abs/2311.12790} {arXiv:2311.12790 [hep-lat]} \BibitemShut {NoStop}%
\bibitem [{\citenamefont {Golterman}\ and\ \citenamefont {Shamir}(2025)}]{Golterman:2025boq}%
  \BibitemOpen
  \bibfield  {author} {\bibinfo {author} {\bibfnamefont {M.}~\bibnamefont {Golterman}}\ and\ \bibinfo {author} {\bibfnamefont {Y.}~\bibnamefont {Shamir}},\ }\bibfield  {title} {\bibinfo {title} {{Constraints on the symmetric mass generation paradigm for lattice chiral gauge theories}},\ }\href@noop {} {\bibfield  {journal} {\bibinfo  {journal} {preprint}\ } (\bibinfo {year} {2025})},\ \Eprint {https://arxiv.org/abs/2505.20436} {arXiv:2505.20436 [hep-lat]} \BibitemShut {NoStop}%
\bibitem [{\citenamefont {Mouland}\ \emph {et~al.}(2025)\citenamefont {Mouland}, \citenamefont {Tong},\ and\ \citenamefont {Zan}}]{Mouland:2025ilu}%
  \BibitemOpen
  \bibfield  {author} {\bibinfo {author} {\bibfnamefont {R.}~\bibnamefont {Mouland}}, \bibinfo {author} {\bibfnamefont {D.}~\bibnamefont {Tong}},\ and\ \bibinfo {author} {\bibfnamefont {B.}~\bibnamefont {Zan}},\ }\bibfield  {title} {\bibinfo {title} {{Phases of 2d Gauge Theories and Symmetric Mass Generation}},\ }\href@noop {} {\bibfield  {journal} {\bibinfo  {journal} {preprint}\ } (\bibinfo {year} {2025})},\ \Eprint {https://arxiv.org/abs/2509.12305} {arXiv:2509.12305 [hep-th]} \BibitemShut {NoStop}%
\bibitem [{\citenamefont {Hasenfratz}\ and\ \citenamefont {Neuhaus}(1989)}]{Hasenfratz:1988vc}%
  \BibitemOpen
  \bibfield  {author} {\bibinfo {author} {\bibfnamefont {A.}~\bibnamefont {Hasenfratz}}\ and\ \bibinfo {author} {\bibfnamefont {T.}~\bibnamefont {Neuhaus}},\ }\bibfield  {title} {\bibinfo {title} {{Nonperturbative Study of the Strongly Coupled Scalar Fermion Model}},\ }\href {https://doi.org/10.1016/0370-2693(89)90899-X} {\bibfield  {journal} {\bibinfo  {journal} {Phys. Lett. B}\ }\textbf {\bibinfo {volume} {220}},\ \bibinfo {pages} {435} (\bibinfo {year} {1989})}\BibitemShut {NoStop}%
\bibitem [{\citenamefont {Sharatchandra}\ \emph {et~al.}(1981)\citenamefont {Sharatchandra}, \citenamefont {Thun},\ and\ \citenamefont {Weisz}}]{Sharatchandra:1981si}%
  \BibitemOpen
  \bibfield  {author} {\bibinfo {author} {\bibfnamefont {H.~S.}\ \bibnamefont {Sharatchandra}}, \bibinfo {author} {\bibfnamefont {H.~J.}\ \bibnamefont {Thun}},\ and\ \bibinfo {author} {\bibfnamefont {P.}~\bibnamefont {Weisz}},\ }\bibfield  {title} {\bibinfo {title} {{Susskind Fermions on a Euclidean Lattice}},\ }\href {https://doi.org/10.1016/0550-3213(81)90200-5} {\bibfield  {journal} {\bibinfo  {journal} {Nucl. Phys. B}\ }\textbf {\bibinfo {volume} {192}},\ \bibinfo {pages} {205} (\bibinfo {year} {1981})}\BibitemShut {NoStop}%
\bibitem [{\citenamefont {Maiti}\ \emph {et~al.}(2026)\citenamefont {Maiti}, \citenamefont {Banerjee}, \citenamefont {Chandrasekharan},\ and\ \citenamefont {Marinkovic}}]{Maiti:2026log}%
  \BibitemOpen
  \bibfield  {author} {\bibinfo {author} {\bibfnamefont {S.}~\bibnamefont {Maiti}}, \bibinfo {author} {\bibfnamefont {D.}~\bibnamefont {Banerjee}}, \bibinfo {author} {\bibfnamefont {S.}~\bibnamefont {Chandrasekharan}},\ and\ \bibinfo {author} {\bibfnamefont {M.~K.}\ \bibnamefont {Marinkovic}},\ }\bibfield  {title} {\bibinfo {title} {{Phase diagram of a lattice fermion model with symmetric mass generation}},\ }\href {https://doi.org/10.1103/tjwn-p6jc} {\bibfield  {journal} {\bibinfo  {journal} {Phys. Rev. D}\ }\textbf {\bibinfo {volume} {114}},\ \bibinfo {pages} {074504} (\bibinfo {year} {2026})},\ \Eprint {https://arxiv.org/abs/2602.18360} {arXiv:2602.18360 [hep-lat]} \BibitemShut {NoStop}%
\bibitem [{\citenamefont {Chandrasekharan}(2010)}]{Chandrasekharan:2009wc}%
  \BibitemOpen
  \bibfield  {author} {\bibinfo {author} {\bibfnamefont {S.}~\bibnamefont {Chandrasekharan}},\ }\bibfield  {title} {\bibinfo {title} {{The Fermion bag approach to lattice field theories}},\ }\href {https://doi.org/10.1103/PhysRevD.82.025007} {\bibfield  {journal} {\bibinfo  {journal} {Phys. Rev. D}\ }\textbf {\bibinfo {volume} {82}},\ \bibinfo {pages} {025007} (\bibinfo {year} {2010})},\ \Eprint {https://arxiv.org/abs/0910.5736} {arXiv:0910.5736 [hep-lat]} \BibitemShut {NoStop}%
\bibitem [{\citenamefont {Zerf}\ \emph {et~al.}(2017)\citenamefont {Zerf}, \citenamefont {Mihaila}, \citenamefont {Marquard}, \citenamefont {Herbut},\ and\ \citenamefont {Scherer}}]{PhysRevD.96.096010}%
  \BibitemOpen
  \bibfield  {author} {\bibinfo {author} {\bibfnamefont {N.}~\bibnamefont {Zerf}}, \bibinfo {author} {\bibfnamefont {L.~N.}\ \bibnamefont {Mihaila}}, \bibinfo {author} {\bibfnamefont {P.}~\bibnamefont {Marquard}}, \bibinfo {author} {\bibfnamefont {I.~F.}\ \bibnamefont {Herbut}},\ and\ \bibinfo {author} {\bibfnamefont {M.~M.}\ \bibnamefont {Scherer}},\ }\bibfield  {title} {\bibinfo {title} {Four-loop critical exponents for the gross-neveu-yukawa models},\ }\href {https://doi.org/10.1103/PhysRevD.96.096010} {\bibfield  {journal} {\bibinfo  {journal} {Phys. Rev. D}\ }\textbf {\bibinfo {volume} {96}},\ \bibinfo {pages} {096010} (\bibinfo {year} {2017})}\BibitemShut {NoStop}%
\bibitem [{\citenamefont {Gehring}\ \emph {et~al.}(2015)\citenamefont {Gehring}, \citenamefont {Gies},\ and\ \citenamefont {Janssen}}]{Gehring2015}%
  \BibitemOpen
  \bibfield  {author} {\bibinfo {author} {\bibfnamefont {F.}~\bibnamefont {Gehring}}, \bibinfo {author} {\bibfnamefont {H.}~\bibnamefont {Gies}},\ and\ \bibinfo {author} {\bibfnamefont {L.}~\bibnamefont {Janssen}},\ }\bibfield  {title} {\bibinfo {title} {Fixed-point structure of low-dimensional relativistic fermion field theories: Universality classes and emergent symmetry},\ }\href {https://doi.org/10.1103/PhysRevD.92.085046} {\bibfield  {journal} {\bibinfo  {journal} {Phys. Rev. D}\ }\textbf {\bibinfo {volume} {92}},\ \bibinfo {pages} {085046} (\bibinfo {year} {2015})}\BibitemShut {NoStop}%
\bibitem [{\citenamefont {Hands}(2021)}]{Hands2021}%
  \BibitemOpen
  \bibfield  {author} {\bibinfo {author} {\bibfnamefont {S.}~\bibnamefont {Hands}},\ }\bibfield  {title} {\bibinfo {title} {{The Planar Thirring Model with K\"ahler-Dirac Fermions}},\ }\bibfield  {journal} {\bibinfo  {journal} {Symmetry}\ }\textbf {\bibinfo {volume} {13}},\ \href {https://doi.org/10.3390/sym13081523} {10.3390/sym13081523} (\bibinfo {year} {2021})\BibitemShut {NoStop}%
\bibitem [{\citenamefont {Hands}\ \emph {et~al.}(1993)\citenamefont {Hands}, \citenamefont {Kocic},\ and\ \citenamefont {Kogut}}]{Hands:1992be}%
  \BibitemOpen
  \bibfield  {author} {\bibinfo {author} {\bibfnamefont {S.}~\bibnamefont {Hands}}, \bibinfo {author} {\bibfnamefont {A.}~\bibnamefont {Kocic}},\ and\ \bibinfo {author} {\bibfnamefont {J.~B.}\ \bibnamefont {Kogut}},\ }\bibfield  {title} {\bibinfo {title} {{Four Fermi theories in fewer than four-dimensions}},\ }\href {https://doi.org/10.1006/aphy.1993.1039} {\bibfield  {journal} {\bibinfo  {journal} {Annals Phys.}\ }\textbf {\bibinfo {volume} {224}},\ \bibinfo {pages} {29} (\bibinfo {year} {1993})},\ \Eprint {https://arxiv.org/abs/hep-lat/9208022} {arXiv:hep-lat/9208022} \BibitemShut {NoStop}%
\bibitem [{\citenamefont {Huang}\ \emph {et~al.}(2025)\citenamefont {Huang}, \citenamefont {Parthenios}, \citenamefont {Ulybyshev}, \citenamefont {Zhang}, \citenamefont {Assaad}, \citenamefont {Classen},\ and\ \citenamefont {Meng}}]{Nature2025-Huang}%
  \BibitemOpen
  \bibfield  {author} {\bibinfo {author} {\bibfnamefont {C.}~\bibnamefont {Huang}}, \bibinfo {author} {\bibfnamefont {N.}~\bibnamefont {Parthenios}}, \bibinfo {author} {\bibfnamefont {M.}~\bibnamefont {Ulybyshev}}, \bibinfo {author} {\bibfnamefont {X.}~\bibnamefont {Zhang}}, \bibinfo {author} {\bibfnamefont {F.~F.}\ \bibnamefont {Assaad}}, \bibinfo {author} {\bibfnamefont {L.}~\bibnamefont {Classen}},\ and\ \bibinfo {author} {\bibfnamefont {Z.~Y.}\ \bibnamefont {Meng}},\ }\bibfield  {title} {\bibinfo {title} {Angle-tuned gross-neveu quantum criticality in twisted bilayer graphene},\ }\href {https://doi.org/10.1038/s41467-025-62461-y} {\bibfield  {journal} {\bibinfo  {journal} {Nature Communications}\ }\textbf {\bibinfo {volume} {16}},\ \bibinfo {pages} {7176} (\bibinfo {year} {2025})}\BibitemShut {NoStop}%
\bibitem [{\citenamefont {Hawashin}\ \emph {et~al.}(2025)\citenamefont {Hawashin}, \citenamefont {Scherer},\ and\ \citenamefont {Janssen}}]{PhysRevB.111.205129}%
  \BibitemOpen
  \bibfield  {author} {\bibinfo {author} {\bibfnamefont {B.}~\bibnamefont {Hawashin}}, \bibinfo {author} {\bibfnamefont {M.~M.}\ \bibnamefont {Scherer}},\ and\ \bibinfo {author} {\bibfnamefont {L.}~\bibnamefont {Janssen}},\ }\bibfield  {title} {\bibinfo {title} {Gross-neveu-$\mathrm{XY}$ quantum criticality in moir\'e dirac materials},\ }\href {https://doi.org/10.1103/PhysRevB.111.205129} {\bibfield  {journal} {\bibinfo  {journal} {Phys. Rev. B}\ }\textbf {\bibinfo {volume} {111}},\ \bibinfo {pages} {205129} (\bibinfo {year} {2025})}\BibitemShut {NoStop}%
\bibitem [{\citenamefont {Campostrini}\ \emph {et~al.}(2001)\citenamefont {Campostrini}, \citenamefont {Hasenbusch}, \citenamefont {Pelissetto}, \citenamefont {Rossi},\ and\ \citenamefont {Vicari}}]{PhysRevB.63.214503}%
  \BibitemOpen
  \bibfield  {author} {\bibinfo {author} {\bibfnamefont {M.}~\bibnamefont {Campostrini}}, \bibinfo {author} {\bibfnamefont {M.}~\bibnamefont {Hasenbusch}}, \bibinfo {author} {\bibfnamefont {A.}~\bibnamefont {Pelissetto}}, \bibinfo {author} {\bibfnamefont {P.}~\bibnamefont {Rossi}},\ and\ \bibinfo {author} {\bibfnamefont {E.}~\bibnamefont {Vicari}},\ }\bibfield  {title} {\bibinfo {title} {Critical behavior of the three-dimensional $\mathrm{XY}$ universality class},\ }\href {https://doi.org/10.1103/PhysRevB.63.214503} {\bibfield  {journal} {\bibinfo  {journal} {Phys. Rev. B}\ }\textbf {\bibinfo {volume} {63}},\ \bibinfo {pages} {214503} (\bibinfo {year} {2001})}\BibitemShut {NoStop}%
\bibitem [{\citenamefont {Butt}\ \emph {et~al.}(2018)\citenamefont {Butt}, \citenamefont {Catterall},\ and\ \citenamefont {Schaich}}]{Butt:2018nkn}%
  \BibitemOpen
  \bibfield  {author} {\bibinfo {author} {\bibfnamefont {N.}~\bibnamefont {Butt}}, \bibinfo {author} {\bibfnamefont {S.}~\bibnamefont {Catterall}},\ and\ \bibinfo {author} {\bibfnamefont {D.}~\bibnamefont {Schaich}},\ }\bibfield  {title} {\bibinfo {title} {{$SO(4)$ invariant Higgs-Yukawa model with reduced staggered fermions}},\ }\href {https://doi.org/10.1103/PhysRevD.98.114514} {\bibfield  {journal} {\bibinfo  {journal} {Phys. Rev. D}\ }\textbf {\bibinfo {volume} {98}},\ \bibinfo {pages} {114514} (\bibinfo {year} {2018})},\ \Eprint {https://arxiv.org/abs/1810.06117} {arXiv:1810.06117 [hep-lat]} \BibitemShut {NoStop}%
\bibitem [{\citenamefont {OpenAI}(2026)}]{openai2025chatgpt}%
  \BibitemOpen
  \bibfield  {author} {\bibinfo {author} {\bibnamefont {OpenAI}},\ }\href {https://openai.com/chatgpt} {\bibinfo {title} {Chatgpt 5.2}} (\bibinfo {year} {2026})\BibitemShut {NoStop}%
\end{thebibliography}%
